\documentclass[12pt]{iopart}

\usepackage{iopams} 
\usepackage{mathrsfs}

\usepackage{graphicx}
\usepackage{amstext}
\usepackage{bm}
\usepackage{amsfonts}
\usepackage{epstopdf}
\usepackage{cite}

\expandafter\let\csname equation*\endcsname=\relax
\expandafter\let\csname endequation*\endcsname=\relax
\usepackage{amsmath,amssymb,amsthm}

\begin{document}

\title[]{Remotely preparing optical Schr\"{o}dinger cat states via homodyne detection in nondegenerate triple-photon spontaneous downconversion}

\author{Miaomiao Wei$^1$, Huatang Tan$^{1,*}$ and Qiongyi He$^{2}$}

\address{$^1$ Department of Physics, Huazhong Normal University, Wuhan 430079, China}
\address{$^2$ State Key Laboratory for Mesoscopic Physics, School of Physics, Frontiers Science Center for Nano-Optoelectronics, and Collaborative Innovation Center of Quantum Matter, Peking University, Beijing 100871, China} 
\address{$^*$ Author to whom any correspondence should be addressed.}

\ead{tht@mail.ccnu.edu.cn and qiongyihe@pku.edu.cn}
\vspace{10pt}

\begin{abstract}
Optical downconversion is a key resource for generating nonclassical states. Very recently, direct nondegenerate triple-photon spontaneous downconversion (NTPSD) with bright photon triplets and strong third-order correlations has been demonstrated in a superconducting device [Phys. Rev. X {\bf 10}, 011011 (2020)]. Besides, linear and nonlinear tripartite entanglement in this process have
 also been predicted [Phys. Rev. Lett {\bf 120}, 043601 (2018); Phys. Rev. Lett. {\bf 125}, 020502 (2020)]. In this paper, we consider the generation of nonclassical optical quantum superpositions and investigate nonlinear quantum steering effects in NTPSD. We find that large-size Schr\"{o}dinger cat states of one downconverted mode can be achieved when the other two modes are subjected to homodyne detection. Also, a two-photon Bell entangled state can be generated when only one mode is homodyned. We further reveal that
such ability of remote state steering originates from nonlinear quantum steerable
correlations among the triplets. This is specifically embodied by the seeming violation of the Heisenberg uncertainty relation for the inferred variances of two noncommutating higher-order quadratures of downconverted modes, based on the outcomes of homodyne
detection on the other mode, i.e., nonlinear quantum steering, compared to
original Einstein-Podolsky-Rosen steering. Our results demonstrate non-Gaussian nonclassical features in NTPSD and would be useful for the fundamental tests of quantum physics and implementations of optical quantum technologies.
\end{abstract}

%
\vspace{2pc}

\section{Introduction}

The generation of Schr\"{o}dinger's cat state \cite{1p} --- a quantum superposition of macroscopically distinct states --- always attracts intense research interests because that it is important not only for understanding fundamentals of quantum mechanics \cite{bas1, bas2,bas3,bas4,bas5, bas6}, such as quantum decoherence \cite{bas4}, but also for a number of quantum technologies \cite{qc1, qc2, qc3, qec, qt1, qt2, qm}.
To date, Schr\"{o}dinger's cat states have been experimentally realized in the systems of e.g. atoms \cite{acat1, acat2, acat3, acat4, acat5} and photons  \cite{acat6, acat7, acat8, acat9, acat10}. In particular, the creation of optical cat states  by photon subtraction on squeezed light has been demonstrated \cite{acat7, acat8, acat9, acat10}, which, however, only generates small-size cat states and moreover depends on single-photon resolved detection. In this situation, when taking into account that homodyne detection has become a relatively mature and high-efficiency technology, it is therefore interesting to generate large-size optical cat states via homodyne detection \cite{he}.

Distinct from second-order twin-photon downconversion which yields Gaussian-state nonclassical light, three-order triple-photon downconversion can create photon triplets in nontrivial non-Gaussian states \cite{tpd00, tpd0, tpd1, tpd2, tpd3, tpd4, tpd5, tpd6, tpd7, tpd8, tpd9, tpd10, tpd11} , e.g., higher-order squeezing \cite{tpd1} which enables error correction in quantum computing \cite{tpd7}.
As we konw, Non-Gaussian states, which may exhibit negative Wigner functions and genuine nonclassicality, has advantageous over Gaussian counterparts in e.g. implementing universal quantum computation \cite{nG}. Spontaneous triple-photon downconversion has been observed in optical fiber but with weak generation rate \cite{3fiber}. Very recently, direct NTPSD with bright photon triplets and strong third-order correlations has been achieved with a superconducting cavity \cite{3sup}. This achievement makes the triplet-photon generation a potential approach to the generation of optical tripartite entanglement and also excites research interests in studying non-Gaussian quantum properties in the process. For instance, three-mode linear-quadrature entanglement in NTPSD \cite{tpd8} was analyzed by Gonz\'{a}lez et al., while Agust\'{\i} et al. found that tripartite genuine non-Gaussian entanglement exists in the process \cite{tpd10}. In addition, some Gaussian and non-Gaussian quantum effects was also revealed by Zhang et al. in Ref. \cite{tpd11}.

In this work, we intend to investigate nonlinear quantum steering effect in NTPSD and its application to the remote generation of optical nonclassical quantum superpositions. Quantum steering reflects the ability of quantum nonlocality to steer
quantum states of a particle which is entangled with another remote
particle subjected to local measurements \cite{st1, st2, st3, st4}. One can exploit quantum steering to remotely prepare desirable quantum states, e.g., quantum quadrature squeezed states \cite{st5,st6}, via homodyne detection.
Quantum steering has been realized in a variety
of systems \cite{gst1, gst2, gst3, gst4, gst5, gst6}, whereas for continuous-variable systems only Gaussian steering has been reported.

We find in this paper that the inferred variances of two noncommutating higher-order quadratures of downconverted modes, conditioned on the linear-quadrature measurements of another distant mode, can seemingly violate
the Heisenberg uncertainty relation, indicating nonlinear quantum steering effect, in comparison
to original Einstein-Podolsky-Rosen steering \cite{st1, epr}. We show that such kind of nonlinear quantum steerable correlations enable the remote generation of large-size single-mode optical Schr\"{o}dinger cat states with high fidelity and two-mode Bell entangled states via homodyne detection. These non-Gaussian optical quantum superpositions exhibit negative Wigner functions, showing strong non-Gaussian nonclassicality. Our results reveal nontrivial non-Gaussian nonclassical features in NTPSD and would be useful for the fundamental tests of quantum physics and implementations of quantum technologies.

\section{System}

We consider a NTPSD process in which a pump photon of high frequency $\omega_p$ is simultaneously down-converted into three low frequency photons, with frequencies $\omega_a$, $\omega_b$ and $\omega_c$ satisfying the energy conservation $\omega_p=\omega_a+\omega_b+\omega_c$.
\begin{figure}[htbp]
\vspace{0.2cm}
\centerline{\hspace{3.8cm}\scalebox{0.52}{\includegraphics{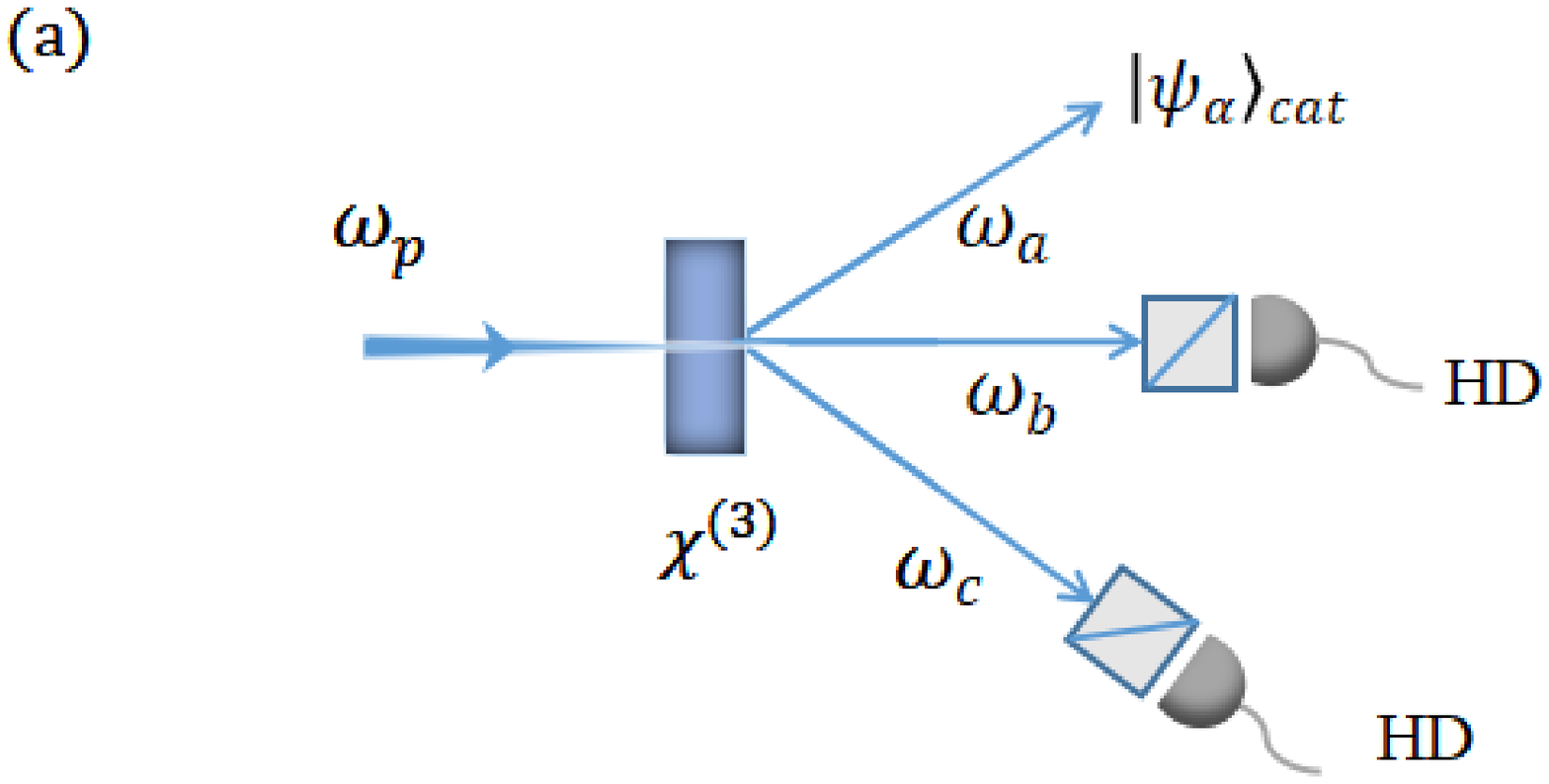}\hspace{2cm}}}
 \vspace{-0.2cm}
 \centerline{\hspace{3.6cm}\scalebox{0.52}{\includegraphics{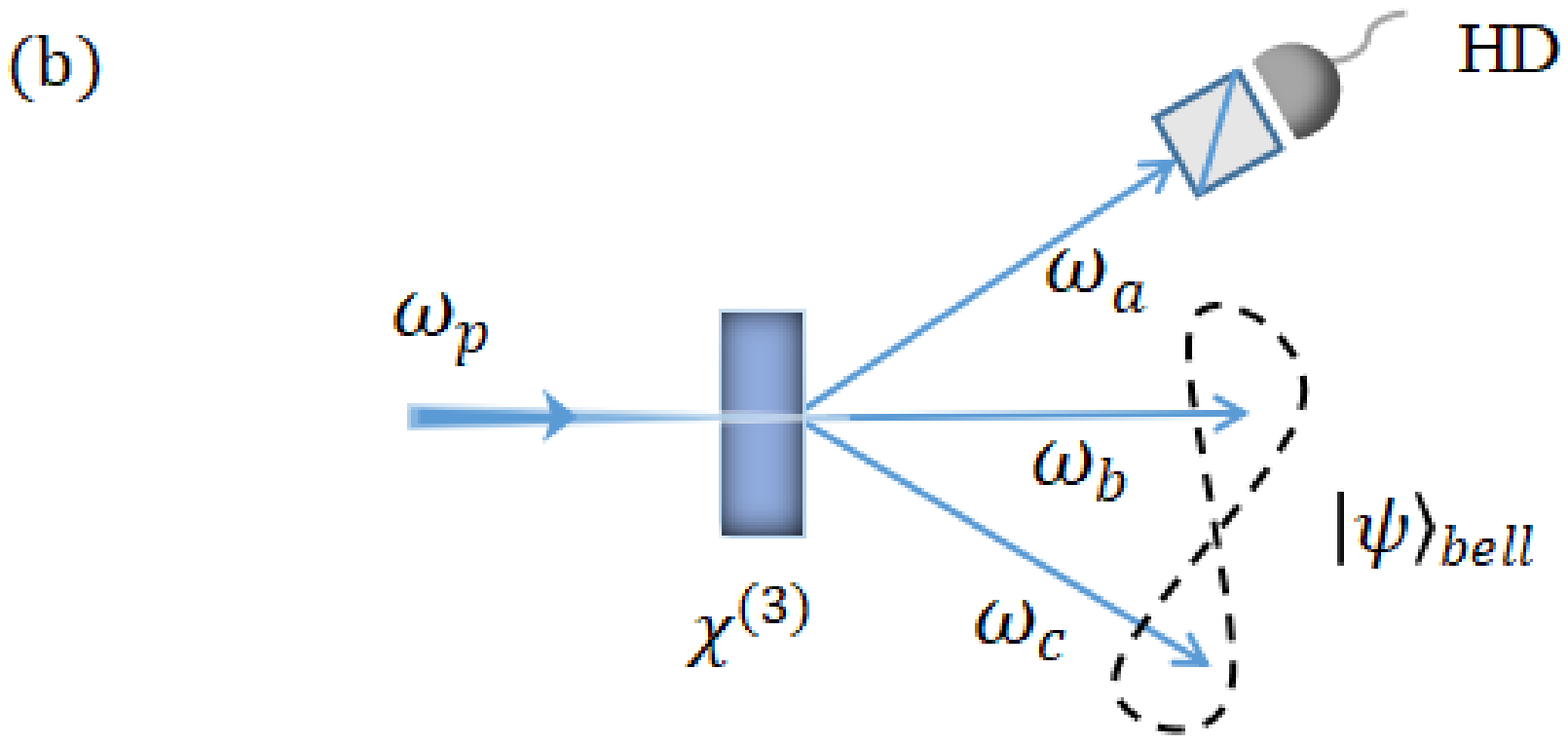}\hspace{2cm}}}
\vspace{-0.5cm}
\caption{Schematic diagram for remotely generating optical quantum superpositions by homodyne detection in a NTPSD process in which a high-frequency pump photon (with frequency $\omega_p$ and in a coherent state $|\alpha_p\rangle$) is simultaneously downconverted into a triplet (denoted by $a,~b,~c$) of frequencies $\omega_a$, $\omega_b$ and $\omega_c$. (a) By homodyning the downconverted modes $b$ and $c$, the third mode $c$ is steered into a size-adjustable Schr\"{o}dinger cat state [$|\psi_\alpha\rangle_{\rm cat}=(|i\alpha \rangle+|-i\alpha\rangle)/\sqrt{2(1+e^{-2|\alpha|^{2}})}]$. (b) When only one mode (e.g. the mode $a$) is homodyned, the other two modes are projected into a two-photon Bell entangled state [$|\psi\rangle_{\rm bell}=(|0_b,0_c\rangle+|1_b,1_c\rangle)/\sqrt{2})$]. The abbreviation ``HD" stands for homodyne detection.}
\label{fig1}
\end{figure}
When taking into account of pump depletion, we can describe this process by using the following Hamiltonian ($\hbar=1$)
\begin{align}
\hat H_q&=i{\chi^{(3)}}(\hat a^\dag\hat b^\dag\hat c^\dag\hat p \ {-} \hat a \hat b\hat c\hat p^\dag),
\end{align}
where $\chi^{(3)}$ is the third-order nonlinear coupling. The annihilation operators $\hat p$, $\hat a$, $\hat b$, and $\hat c$ and their conjugates denote the pump mode and three down-converted modes, respectively.
Experiments have realized optical NTPSD in optical fiber \cite{3fiber}, with small triplet generation rate. While in the microwave domain, Chang et al. have very recently achieved direct NTPSD by using a flux-pumped superconducting parametric cavity and also observed strong three-body correlations among the triple modes \cite{3sup}.

Given initial vacua of the triplets $|0_a,0_b,0_c\rangle$ and a coherent state of the pump mode $|\alpha_{p}\rangle=e^{-|\alpha_p|^2/2}\sum_{n_p}\alpha_p^n|n_p\rangle/\sqrt{n_p!}$, with the amplitude $\alpha_p$, the state's evolution of the whole system is governed by
\begin{align}
\frac{d|\psi(t)\rangle_{abcp}}{dt}=-i\hat H_q |\psi(t)\rangle_{abcp},
\end{align}
By expanding the state vector in the Fock space $\{|n_a,n_b,n_c,n_p\rangle\}$, $|\psi(t)\rangle_{abcp}=\sum_n c_{n,n_p}(t)|n_a,n_b,n_c,n_p-n\rangle$, with $n_{a, b, c}=n$ for the photon-triplet generation, the amplitude $c_{n,n_p}(t)$ and thus the state vector can be obtained numerically. Since we are only interested the photon triplets, the reduced state $\hat{\rho}_{abc}(t)$ of the triplets can be obtained by tracing out the pump mode.

\section{Optical quantum superpositions} 
With the density matrix $\hat \rho_{abc}$ of the triplets, we can study remotely steering quantum states of downconverted modes via homodyne detection on the linear quadratures $\hat X_o=(\hat o+\hat o^\dag)/\sqrt{2}~(o=a,b,c)$ of the other modes. We first investigate the state of the mode $a$ when simultaneously homodyning the quadratures $\hat X_b$ and $\hat X_c$. Conditioned on the measurement outcomes $x_b$ and $x_c$, the density operator $\hat \rho_a^{\rm con} (x_b, x_c)$ can be formally given by
\begin{align}
\hat \rho_a^{\rm con}(x_b,x_c)=\frac{\hat {\tilde{\rho}}_a^{\rm con}(x_b,x_c)}{\text{Tr}_a \big[\hat {\tilde{\rho}}_a^{\rm con}(x_b,x_c)\big]},
\end{align}
where $\hat {\tilde{\rho}}_a^{\rm con}(x_b,x_c)=\text{Tr}_{bc} \big[(\hat {\mathcal M}_{bc} \otimes \hat I_a)\hat \rho_{abc}(\hat I_a\otimes\hat {\mathcal M}_{bc})\big]$ and $\hat {\mathcal M}_{bc}=|x_b,x_c\rangle\langle x_b,x_c|$ which can be calculated in the Fock space with $\langle x_o\mid n_o\rangle=\frac{1}{\pi^{1/4}}\frac{1}{\sqrt{2^{n_o}n_o!}}e^{-x_{o}^{2}/2}H_{n_o}(x_o)$,
$H_{n}$ the Hermite polynomial of order $n$. The superscript ``con" stands for ``conditional" and similarly hereinafter.

\begin{figure}[t]
\vspace{0cm}
\centerline{\hspace{2cm}\scalebox{0.5}{\includegraphics{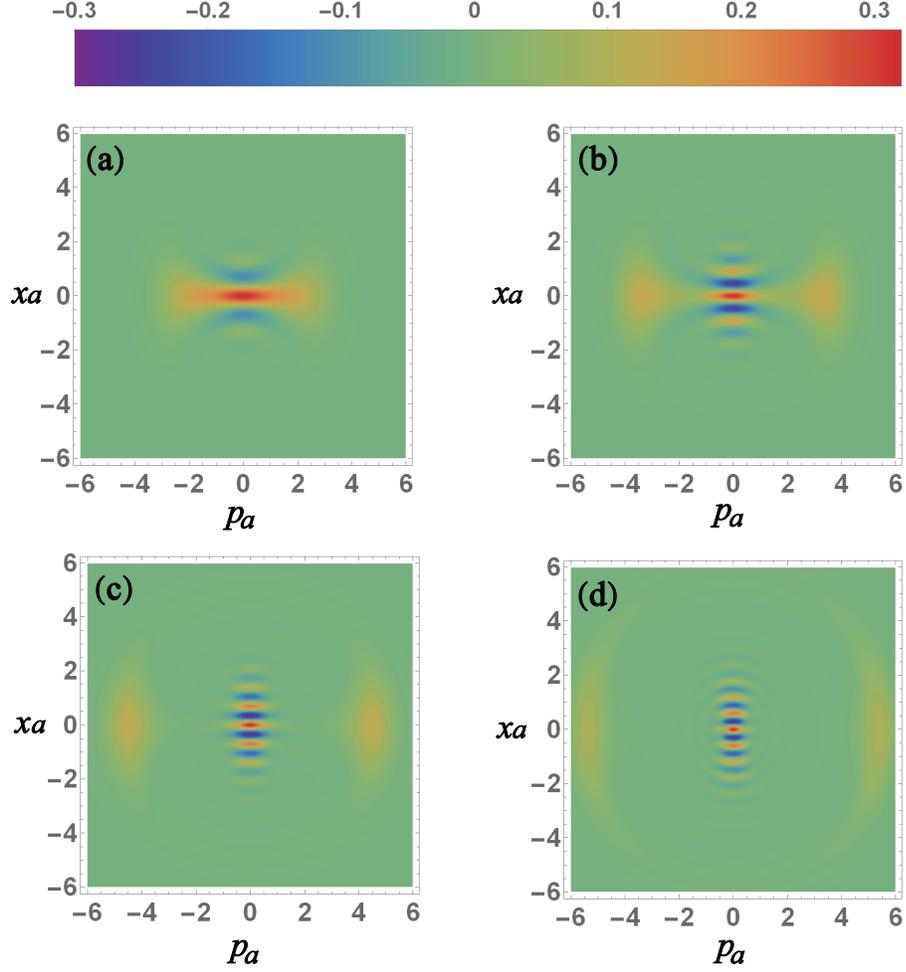}}}
\caption{The density plots of the Wigner function $W_a^{\rm con}(x_a, p_a)$  for the measurement results $x_c=0$ and (a) $x_b=3$, (b) $x_b=4$, (c) $x_b=5$, and (d) $x_b=6$. The interaction time $\alpha_{p}\chi^{(3)} t=0.3$ and pump amplitude $\alpha_{p}=\sqrt{10}$.}
\label{fig2}
\end{figure}

We consider the Wigner function $W_a^{\rm con}(x_a,p_a)$ of the density matrix $\hat \rho_a^{\rm con}(x_b,x_c)$, obtained by performing Fourier transform on the characteristic function defined via $\chi_a(\xi)=\text{Tr}\big[e^{\xi\hat a^\dag -\xi^*\hat a}\hat \rho_a^{\rm con}(x_b,x_c)\big]$. In Fig.2, the density plots of the Wigner function $W_a^{\rm con}(x_a,p_a)$ for the measurement outcomes $x_c=0$ and $x_b=\{3,4,5,6\}$ are presented at the interaction time $\alpha_p\chi^{(3)}t=0.3$. It shows that the Wigner function exhibits obvious negativity, indicating strong non-Gaussian nonclassicality. In Fig.3~(a), we plot the
the fidelity $F_{a}=\text{Tr}\big[\sqrt{\sqrt{\hat \rho_\alpha}\hat \rho_a^{\rm con}\sqrt{\hat \rho_\alpha}}\big]$, with respect to the ideal cat state $\hat \rho_\alpha=|\psi_\alpha\rangle_{\rm cat}\langle \psi_\alpha|$ and
\begin{align}
|\psi_\alpha\rangle_{\rm cat}=\frac{1}{\sqrt{2(1+e^{-2|\alpha|^{2}})}}\big(|i\alpha \rangle+|-i\alpha\rangle\big).
\label{cat}
\end{align}
We see that high fidelity can be achieved. For instance, $F_{a}\approx 0.98$ for the amplitude $\alpha\approx 1.2$. The downconverted mode $a$ is thus steered by the measurement into a cat state remotely. Further, the size of the cat states can be adjustable and, as shown explicitly, it grows up but still with high fidelity when the measurement outcome increases from $x_b=3$ to $x_b=6$. For example, high-fidelity cat state ($F_a\approx 0.95$) with large amplitude $\alpha\approx 3.2$ can be achieved. We see from Fig.3 (b) the purity of the state $\hat \rho_a^{\rm con}$, defined by $P_a=\rm Tr[(\hat \rho_\emph{a}^{\rm con})^2]$, is almost equal to unit and just slightly decreases as $x_b$ increases from $x_b=3$. Fig. 2 also shows that the two peaks in $p_a$ become more separated with larger values of the outcome $x_b$, representing larger size of the created cats but still with stronger nonclassicality $\mathcal N_a$ and macroscopicity $\mathcal M_a$, as shown in Fig.3~(b), which defined, respectively, by \cite{noncla}
\begin{align}
\mathcal N_a\equiv\int\big(\big|W(\beta,\beta^{*})\big|-W(\beta,\beta^{*})\big)d^{2}\beta,
\end{align}
and \cite{macro}
\begin{align}
\mathcal M_a\equiv\frac{\pi}{2}\int W(\beta,\beta^{*})\big(-\frac{\partial^{2}}{\partial\beta\partial\beta^{*}}-1\big)
W(\beta,\beta^{*})d^{2}\beta.
\end{align}
We see that the nonclassicality has a peak around $x_b\approx 5 $ since the optimal fidelity decreases as $x_b$ increases. The macroscopicity still increases with the increasing of $x_b$. In addition, as shown from the time evolution of the optimal fidelity $F_{a}^{\rm max}$, with respect to the cat-state amplitude $\alpha$, and the negativity $\mathcal N_a$ in Fig.4, for the fixed $x_b=3$ the negativity becomes saturated with time, although the optimal fidelity deceases, since the amplitude $\alpha$ increases. At time $\alpha_p\chi^{(3)}t=1$, $F_{a}^{\rm max}\approx0.83$, with respect to the cat state of amplitude $\alpha=2.2$, and $\mathcal N_a\approx0.58$.
Therefore, the adjustable and large-amplitude optical cat states can be generated by homodyne detection for the present scheme. We note that small-size cat states (kitten) (the corresponding amplitude $\alpha \lesssim 1.6$) can only be realized via photon-subtraction detection \cite{acat8, acat9, acat10}.

\begin{figure}[t]
\centerline{\scalebox{0.28}{\hspace{8cm}\includegraphics{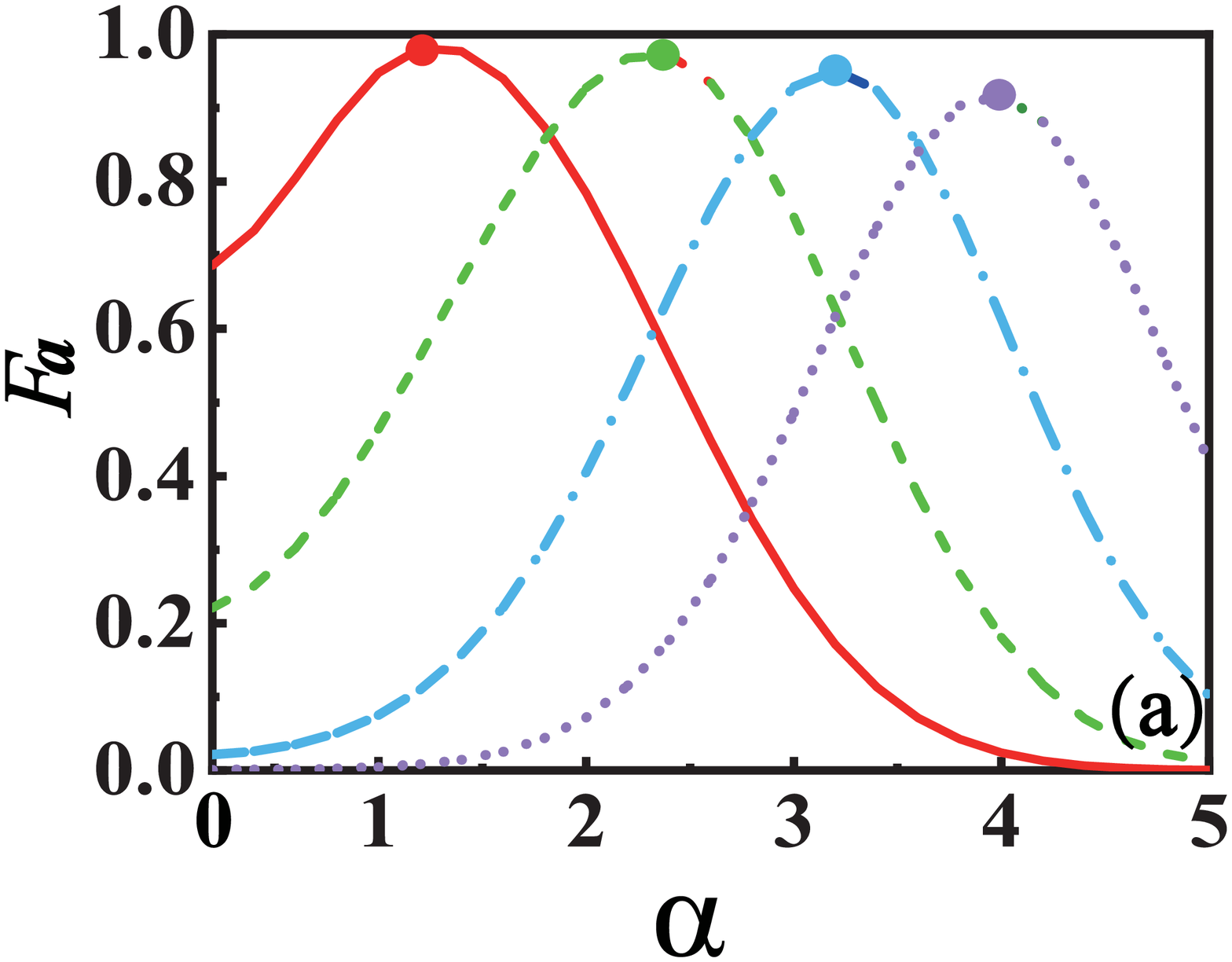}}\scalebox{0.3}{\hspace{0.3cm}\includegraphics{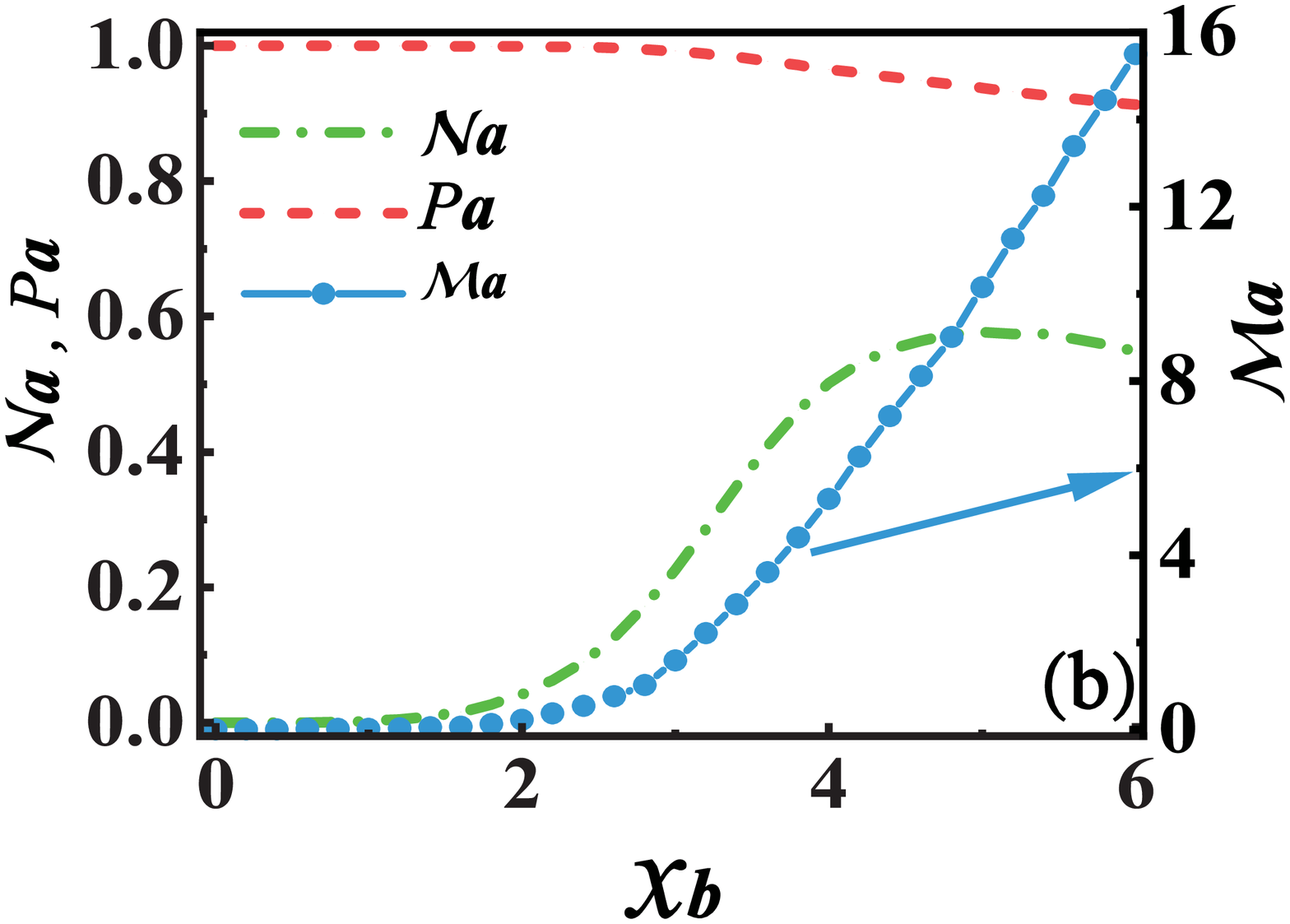}}}
\caption{(a) The fidelity $F_{a}$ between the achieved state $\hat \rho_{a}^{\rm con}$ and the ideal cat state $|\psi _\alpha\rangle$ as the function of the cat-state amplitude $\alpha$, for different measurement results $x_c=0$, $x_b=3~(red)$, $4~(green)$, $5~(blue)$, and $6~(purple)$.  The fidelity  and amplitude, represented by the marked points, are respectively given by $\{F_a, \alpha\}\approx\{0.98, 1.2\}$, $\{0.97, 2.4\}$, $\{0.95,3.2\}$, and $\{0.92, 4\}$. (b) The dependence of the Wigner negativity $\mathcal N_a$, the purity $P_{a}$ and macroscopic quantum superposition $\mathcal M_a$ of the state $\hat \rho_{a}^{\rm con}$ on the measurement outcomes $x_{b}$, for $x_{c}$=0. The interaction time $\alpha_{p}\chi^{(3)} t=0.3$ and pump amplitude $\alpha_{p}=\sqrt{10}$.}
\label{fig3}
\end{figure}

\begin{figure}[t]
\centerline{\scalebox{0.28}{\hspace{6cm}\includegraphics{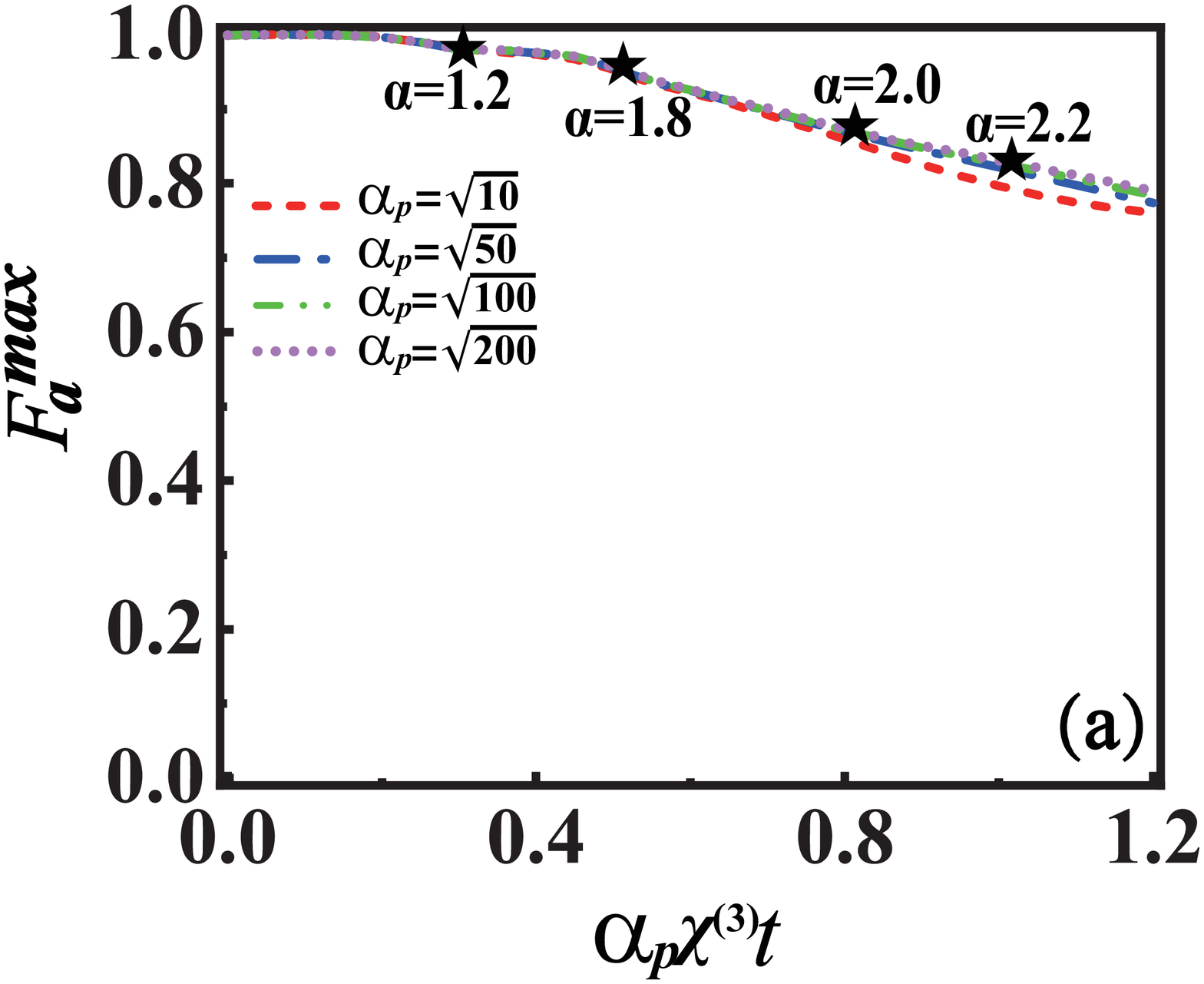}}\scalebox{0.276}{\hspace{0cm}\includegraphics{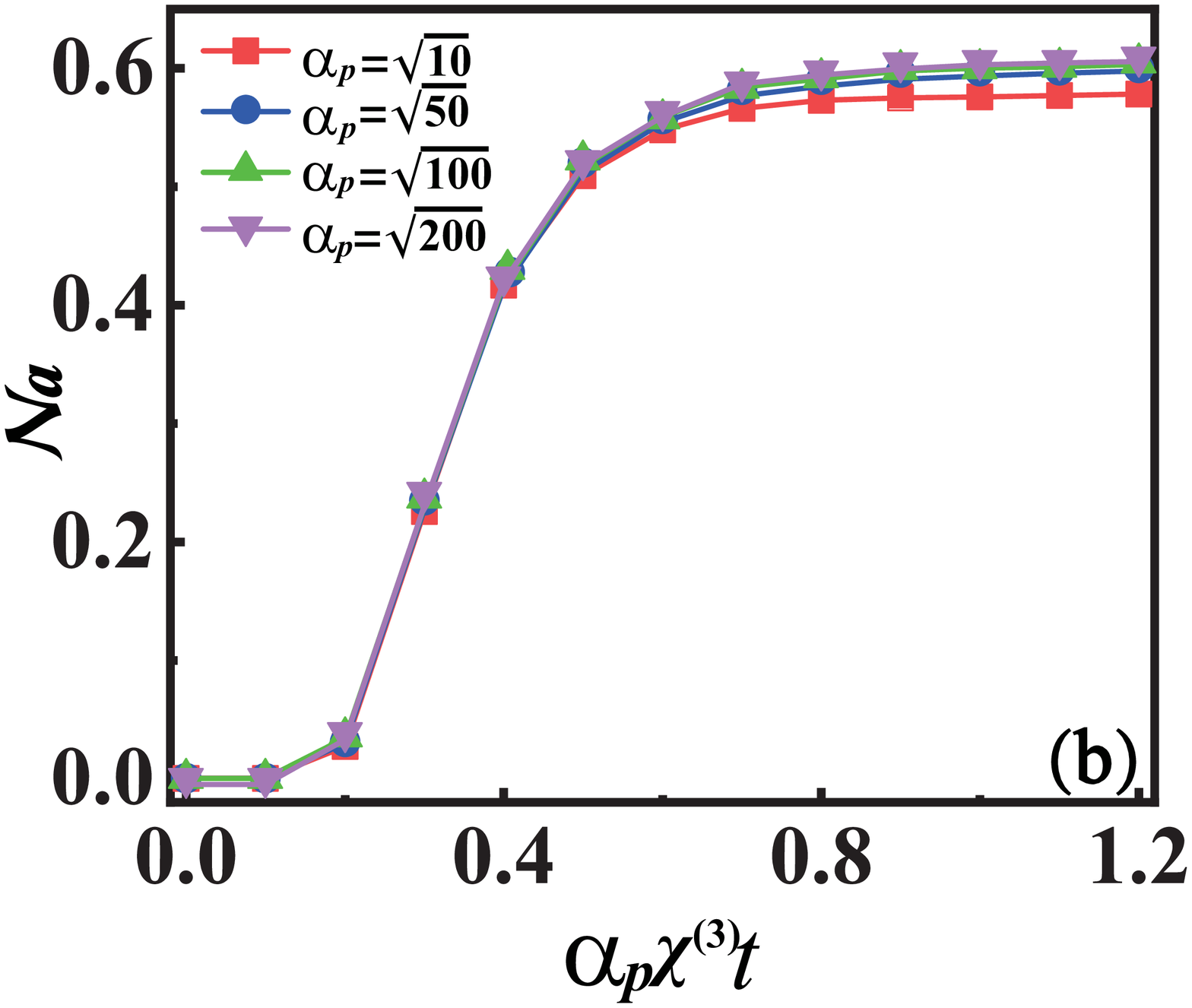}}}
\caption{The time dependences of the optimal fidelity $F_{a}^{\rm max}$ over different cat-state amplitudes $\alpha$ in (a) and the negativity $\mathcal N_a$ of the state $\hat \rho_{a}^{\rm con}$ in (b), for fixed measurement outcomes $x_{b}=3$ and $x_{c}$=0. The marked stars denote the optimal fidelity and the amplitude $\{F_a^{\rm max},\alpha\}\approx\{0.98, 1.2\}$, $\{0.96,1.8\}$, $\{0.87,2.0\}$ and $\{0.83,2.2\}$.}
\label{fig4}
\end{figure}

We next study the two-mode state ${\hat{\rho}}_{bc}^{\rm con}(x_a)$ of the modes $b$ and $c$ generated by homodying on the quadrature $\hat X_a$ of the mode $a$. In Fig.4~(a), we plot the fidelity $F_{bc}$ of ${\hat{\rho}}_{bc}^{\rm con}(x_a)$ with respect to a maximally entangled Bell state
\begin{align}
|\psi\rangle_{\rm bell}=\frac{1}{\sqrt{2}}\big(|0_b,0_c\rangle+|1_b,1_c\rangle\big).
\end{align}

\begin{figure}[htbp]
\centerline{\scalebox{0.268}{\hspace{6cm}\includegraphics{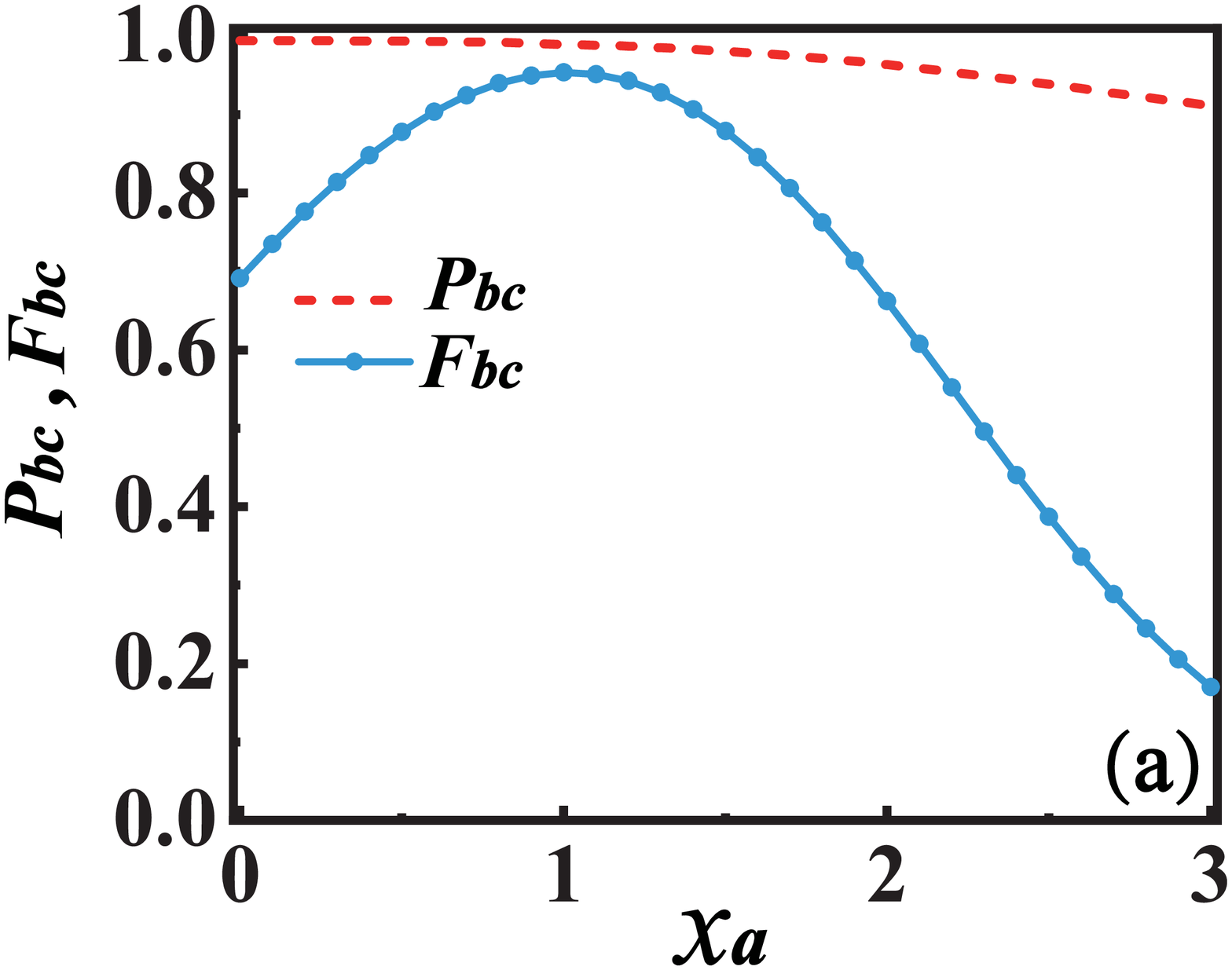}}\scalebox{0.288}{\hspace{0.5cm}\includegraphics{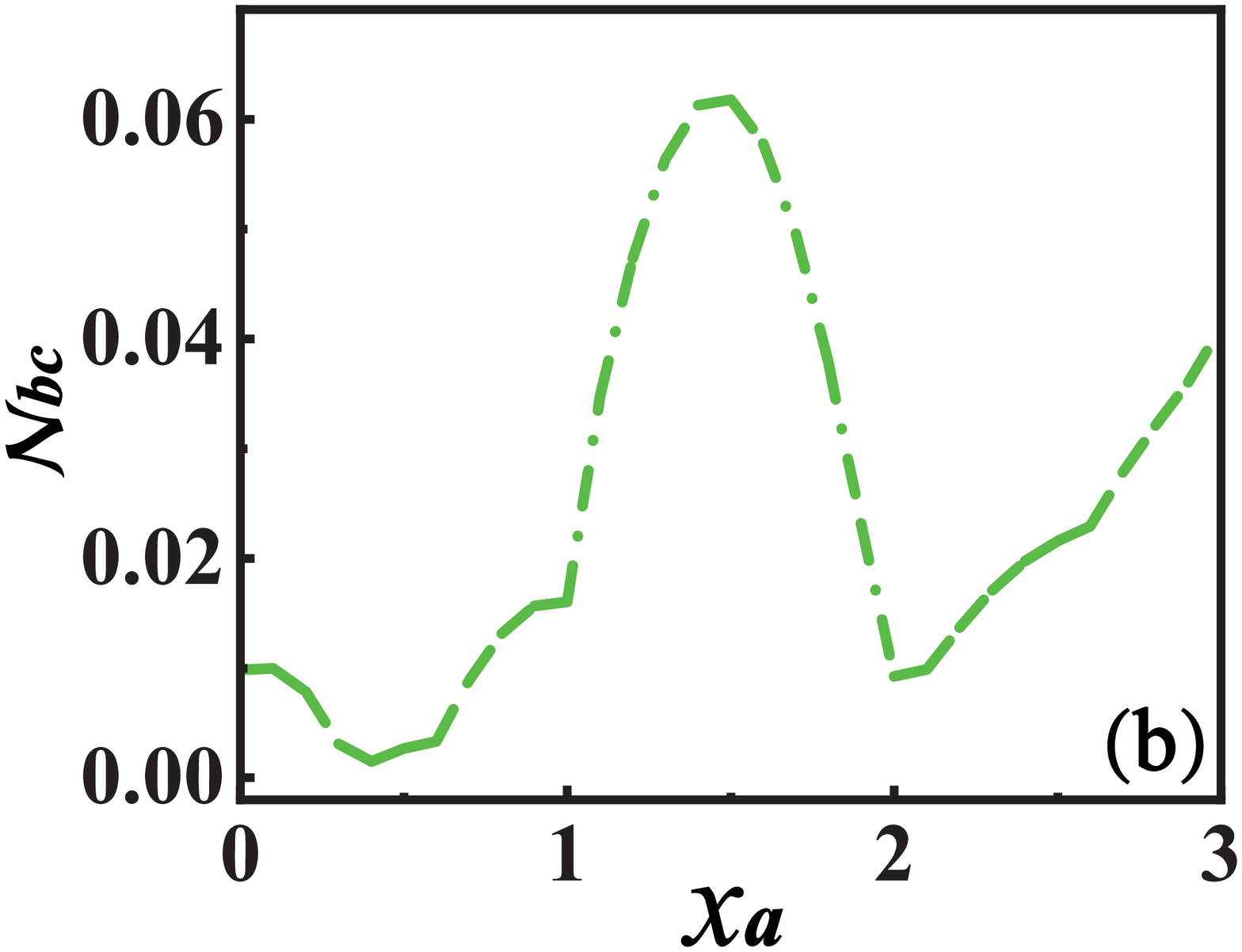}}}
\caption{The purity $P_{bc}$ and fidelity $F_{bc}$ in (a) and the Wigner-function negativity $\mathcal {N}_{bc}$ in (b) of the density matrix ${\hat{\rho}}_{bc}^{\rm con}(x_a)$ as the function of  the measurement result $x_{a}$. The interaction time $\alpha_{p}\chi^{(3)} t=0.6$ and $\alpha_{p}=\sqrt{10}$.}
\label{fig5}
\end{figure}

The optimal fidelity $F_{bc}\approx 0.95$ around the outcome $x_a=1$ at which the purity $P_{bc}\approx 0.99$ of the density matrix ${\hat{\rho}}_{bc}^{\rm con}$, as shown in Fig.5 (a). Therefore, an approximate Bell entangled pure state $|\psi\rangle_{bc}^{\rm con}\approx0.817|0_b,0_c\rangle+0.531|1_b,1_c\rangle$ is generated via homodyne detection. Such qubit entanglement is essential for scalable photonic quantum devices \cite{qbt}. In Fig.5~(b), the negativity $\mathcal {N}_{bc}$ of the Wigner function of ${\hat{\rho}}_{bc}^{\rm con}(x_a)$ is plotted. It is shown the negativity is present, indicating genuine non-Gaussian nonclassicality. In addition, we see that the maximal negativity occurs in the vicinity of $x_a=1.5$ at which the fidelity is not optimal. This is because that the probabilities of single- and two-photon states in the modes $a$ and $b$ increase as $x_a$ arises from 1 to 1.5. At $x_a=1.5$, the state becomes into a qudit entangle state: $|\psi\rangle_{bc}^{\rm con}\approx0.63|0_b,0_c\rangle+0.613|1_b,1_c\rangle)+0.423|2_b,2_c\rangle$, studied in Ref.\cite{qdt}.

\section{Nonlinear quantum steering}
The above results show that the measurements of normal linear quadratures of one or two modes can project the other modes into non-Gaussian nonclassical states. This is in fact due to the nonclassical correlations built up in the NTPSD process since the triplet's state $|\psi\rangle_{abc}(t)$ in the short-time regime $
|\psi\rangle_{abc}(t)\approx |0_a,0_b,0_c\rangle+gt|1_a, 1_b, 1_c\rangle+\frac{g^2t^2}{2}|2_a, 2_b, 2_c\rangle$,
according to the Hamiltonian $\hat H_c=ig(\hat a^\dag\hat b^\dag\hat c^\dag \ - \hat a \hat b\hat c)$, for strong driving such that pump depletion can be neglected, where $g=\chi^{(3)}\bar \alpha_p$ with real pump amplitude $\bar \alpha_p$. Specifically, the correlations are quantum steerable, allowing us to steer desirable quantum states via remote measurement, in the spirit of the original EPR steering \cite{st1}. However, different from the latter which is linear and merely leads to conditional Gaussian states squeezing via homodyne detection \cite{st4}, they are nonlinear. To reveal the nonlinear steering effects, let us at first define the generalized higher-order quadratures
\begin{align}
\hat{X}_{bc}=\frac{\hat{b}\hat{c} \ {+} \hat{b}^\dag\hat{c}^\dag}{2}, ~~~~ \hat{Y}_{bc}=\frac{\hat{b}\hat{c} \ {-} \hat{b}^\dag\hat{c}^\dag}{2i}.
\end{align}
The steering from the mode $a$ to the group of the modes $b$ and $c$ can be characterized by the product of the inferred variances of the quadratures $\hat{X}_{bc}$ and $\hat{Y}_{bc}$ in the two-mode state $\hat {\rho}_{bc}^{\rm con}$ from the measurements of the normal quadratures $\hat X_a$ and $\hat Y_a$ of the mode $a$ \cite{st2}, i.e.,
\begin{align}
S^{a\rightarrow bc}=\frac{2\sqrt{\Big\langle \Delta\big(\hat X_{ bc}| x_a\big)\Big\rangle_e\Big\langle\Delta\big(\hat Y_{bc}|y_a\big)\Big\rangle_e}}{ \Big\langle\big[\hat X_{bc},\hat Y_{bc}\big]\Big\rangle_e}<1.
\label{str}
\end{align}

\begin{figure}[t]
\centerline{\scalebox{0.63}{\hspace{3cm}\includegraphics{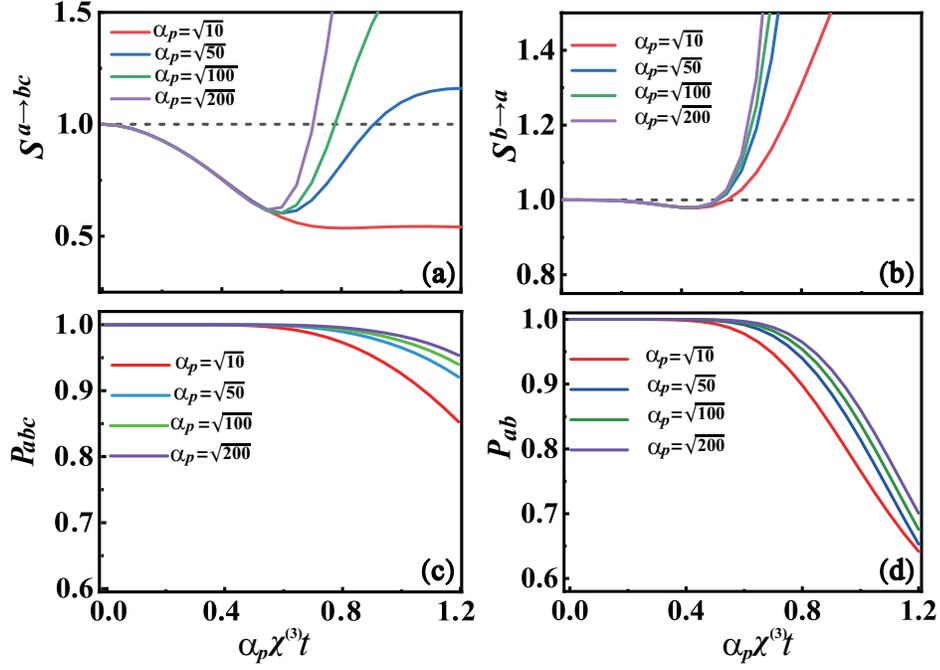}}}
\caption{Time evolution of steering quantities $S^{a\rightarrow bc}$ in (a), $S^{b\rightarrow a}$, the purities $P_{abc}$ and $P_{ab}$ of the reduced state $\hat{\rho}_{abc}$ in (c) and $\hat{\rho}_{ab}^{\rm con}$  with  $x_c=0$ in (d). The pump coherent-state amplitudes $\alpha_{p}=\sqrt{10},~\sqrt{50},~\sqrt{100}$, and $\sqrt{200}$.}
\label{fig6}
\end{figure}

Here $\Delta\big(\hat {\mathcal O}|r\big)\equiv\text{T}r \big[\hat {\rho}_{bc}^{\rm con}(r)\hat {\mathcal O}^2\big]-\Big[\text{T}r \big[\hat {\rho}_{bc}^{\rm con}(r)\hat {\mathcal O}\big]\Big]^2$ is the inferred variance of the operator $\hat {\mathcal O}=\big\{\hat X_{bc},\hat Y_{bc}\big\}$ for the two-mode states $\hat \rho_{bc}^{\rm con}(r=x_a)$ and $\hat \rho_{bc}^{\rm con}(r=y_a)$ conditioned on the outcomes $r=x_a$ and $r=y_a$ of the corresponding homodyne detection on $\hat X_a$ and $\hat Y_a$, respectively. The symbol $\langle \cdot \rangle_e$ denotes the ensemble average over all possible outcomes $x_a$ and $y_a$. We have, for instance, $\big\langle\Delta\big(\hat {\mathcal O}|r\big)\big\rangle_e=\int dr P(r)\Delta\big(\hat {\mathcal O}|r\big)$, with the probability distribution $P(r)$, and similarly for the others. It can be seen from Eq.(9) that the higher-order variances seemingly violates the Heisenberg uncertainty relation, embodying the essence of the original EPR paradox \cite{epr}. Note that this is just a sufficient measure for the existence of nonlinear quantum steering. In addition, it can be implied from Eq.(9) that
\begin{align}
\frac{\Big\langle \Delta\big(\hat X_{ bc}| x_a\big)\Big\rangle_e}{\Big\langle\big[\hat X_{bc},\hat Y_{bc}\big]\Big\rangle_e}<\frac{1}{2}~~~\text{or/and}~~~\frac{\Big\langle \Delta\big(\hat Y_{ bc}| y_a\big)\Big\rangle_e}{\Big\langle\big[\hat X_{bc},\hat Y_{bc}\big]\Big\rangle_e}<\frac{1}{2},
\label{nsq}
\end{align}
showing that the nonlinear steering also means the presence of unconditional (or conditional) higher-order squeezing of the  operators $\hat X_{bc}$ or $\hat Y_{bc}$.

In Fig.6~(a) and (c), the time evolution of the steering quantity $S^{a\rightarrow bc}$ and the purity $P_{abc}\equiv \text{Tr}(\hat \rho_{abc}^2)$ of the reduced state $\hat \rho_{abc}$ is plotted for different pump coherent-state amplitude $\alpha_{p}$. We see $S^{a\rightarrow bc}<$1 in the short-time regime, showing nonlinear quantum steering from the mode $a$ to the group of the modes $b$ and $c$.   Due to the symmetry among the triplets, it can therefore be concluded that the tripartite nonlinear steering can be built up among the triplets via the NTPSD process. As the amplitude $\alpha_p$ decreases, it takes longer time for $S^{a\rightarrow bc}$ to reach its minimal value (stronger steering), since the effective coupling $\chi^{(3)}\alpha_p$ decreases and it requires more time for accumulation. Also, the minima of $S^{a\rightarrow bc}$ become larger (weaker steering) as $\alpha_p$ increases, because of larger Fock space $\{|n_p\rangle\}$ occupied by the pump state, leading to less amount of nonclassicality. We note that in the experiment \cite{3sup} strong coherent pump was employed to ensure large generation rate of the triplets. At the beginning of time development, the tripartite state is almost pure and it becomes mixed as the time continuously develops. The purity increases as the amplitude $\alpha_p$ increases, since this is closer to the case of classical treatment of the pump.

We further investigate the nonlinear steering between the two modes $a$ and $b$ in the conditional state $\hat \rho_{ab}^{\rm con}(x_c)$ which is dependent on the outcome $x_c$. Similarly, the squared amplitudes of the mode $a$ are defined by
\begin{align}
\hat{X}_{a^2}=\frac{\hat{a}^2 \ {+} \hat{a}^{\dag2}}{2},~~~~
\hat{Y}_{a^2}=\frac{\hat{a}^2 \ {-} \hat{a}^{\dag2}}{2i}.
\end{align}
We only consider the steering from the mode $b$ to the mode $a$ via homodying the normal linear quadrature $\hat X_b$ and $\hat Y_b$, quantified by $S^{b\rightarrow a}$ which is defined in the same way as Eq.(\ref{str}) just by replacing the operators $\big\{\hat X_{bc},\hat Y_{bc}\big\}\rightarrow \big\{\hat X_{a^2},\hat Y_{a^2}\big\}$, with the variances of $\hat X_{a^2}$ and $\hat Y_{a^2}$ respectively in the states $\hat \rho_a^{\rm con}(x_b,x_c)$ and $\hat \rho_a^{\rm con} (x_b, y_c)$ and the ensemble average over the outcomes $x_b$ and $y_b$.

In Fig.6 (b) and (d), the quantity $S^{b\rightarrow a}$ and the purity $P_{ab}$ of the conditional state $\hat \rho_{ab}^{\rm con}$ are respectively plotted for $x_c=0$. It is shown that the steering can also exist in the transient regime but the degree of the steering is weaker than that of $S^{a\rightarrow bc}$. The purity of the conditional state is higher than that of the unconditional state $\hat \rho_{abc}$, due to the measurement enhancing the purity.

\begin{figure}[htbp]
\centerline{\scalebox{0.62}{\hspace{3cm}\includegraphics{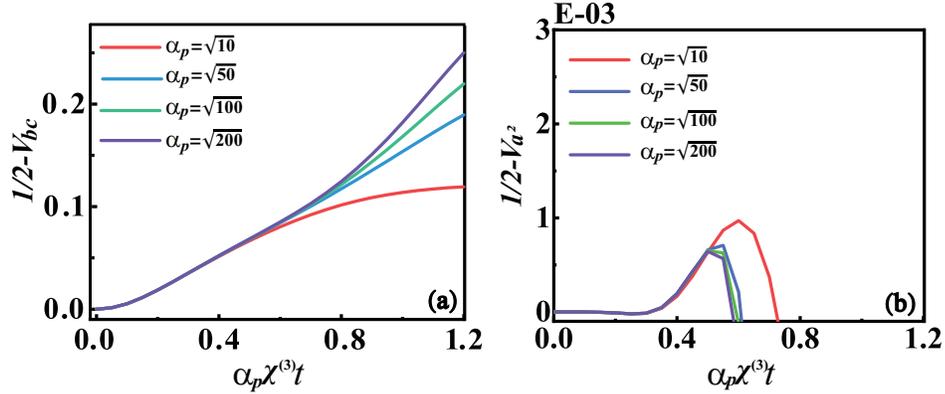}}}
\caption{Time evolution of the conditional higher-order squeezing $V_{bc}$ and $V_{a^2}$ of the quadratures $\hat X_{bc}$ for $x_a=0$ in (a) and $\hat X_{a^2}$ for both $x_b=3$ and $x_c=0$ in (b).}
\label{fig7}
\end{figure}

Finally, in Fig.7~(a) and (b) we plot the quantities 
\begin{align}
 V_{bc}=\frac{\Delta\big(\hat X_{ bc}| x_a\big)}{\big[\hat X_{bc},\hat Y_{bc}\big]} , ~~~~V_{a^2}=\frac{\Delta\big(\hat X_{a^2}| (x_b,x_c)\big)}{\big[\hat X_{a^2},\hat Y_{a^2}\big]}. 
\end{align}
which depict the conditional higher-order squeezing of the quadratures $\hat X_{bc}$ and $\hat X_{a^2}$ when $V_{bc}<1/2$ and $V_{a^2}<1/2$. We see that compared to the nonlinear steering, the conditional higher-order squeezing  exist in the longer-time regime because that the ensemble average in Eq.(\ref{nsq}) is not performed. The squeezing of $\hat X_{bc}$ is much larger than $\hat X_{a^2}$, since the former corresponds to stronger quantum steering.

In summary, we have studied the generation of  optical quantum superpositions and nonlinear quantum steering effect in NTPSD. It is found that the remote generation of large-size single-mode optical Schr\"{o}dinger cat states with high fidelity and two-mode Bell entangled states can be achieved by homodyne detection. This is enabled by the nonlinear steerable correlations among triplets, embodying by seeming violation of
the Heisenberg uncertainty relation of the inferred variances of two noncommutating higher-order quadratures of downconverted modes, conditioned on the linear-quadrature measurements of another distant mode.  Our results reveal non-Gaussian nonclassical features in NTPSD and would be useful for the fundamental tests of quantum physics and implementations of optical quantum tasks. Further study will include the investigation of nonclassical features of output fields for an intracavity NTPSD in the pulse and continuous pump regimes \cite{out}.

\ack

  This work is supported by the National Natural Science Foundation of China(Nos.11674120 and 12174140) and the Fundamental Research Funds for the Central Universities (CCNU20TD003). Q.Y.He is supported by the National Natural Science Foundation of China (Grants No. 61675007, No. 11975026, and No. 61974067), the Key R\&D Program of the Guangzhou Province (Grant No. 2018B030329001), and the Beijing Natural Science Foundation (Grant No. Z190005).

\section*{Data availability statement}
The data that support the findings of this study are available upon reasonable request from the authors.

\end{document}